\documentstyle[12pt]{article}
\def\gev{\; {\rm GeV}}

\def\bbar{\overline{\beta}}
\def\mbar{\overline{m}}
\def\veff{V_{eff}(\phi,T)}

\def\v1t{V^{(1)}[m^2(\phi),T]}

\def\ov{\overline}
\def\simlt{\stackrel{<}{{}_\sim}}
\def\simgt{\stackrel{>}{{}_\sim}}
\def\to{\rightarrow}
\newcommand{\be}{\begin{equation}}
\newcommand{\ee}{\end{equation}}
\begin{document}
\begin{titlepage}
\phantom{bla}
\hfill{CERN-TH.6577/92}
\\
\phantom{bla}
\hfill{IEM-FT-58/92}
\vskip 2.5cm
\begin{center}
{\Large\bf On the nature of the electroweak phase transition}
\end{center}
\vskip 1.5cm
\begin{center}
{\large J. R. Espinosa} \footnote{Supported by a grant of Comunidad de
Madrid, Spain.},
{\large M. Quir\'os} \footnote{Work partly supported by CICYT, Spain,
under contract AEN90-0139.}
\\
\vskip .3cm
Instituto de Estructura de la Materia, CSIC\\
Serrano 123, E-28006 Madrid, Spain\\
\vskip .3cm
and \\
\vskip .3cm
{\large F. Zwirner} \footnote{On leave from INFN, Sezione di Padova, Padua,
Italy.}
\\
\vskip .3cm
Theory Division, CERN \\
CH-1211 Geneva 23, Switzerland \\
\end{center}
\vskip 1cm
\begin{abstract}
\noindent
We discuss the finite-temperature effective potential of the Standard Model,
$\veff$, with emphasis on the resummation of the most important infrared
contributions. We compute the one-loop scalar and vector boson self-energies in
the zero-momentum limit. By solving the corresponding set of gap equations,
with the inclusion of subleading contributions, we find a non-vanishing
magnetic mass for the $SU(2)$ gauge bosons. We comment on its possible
implications for the nature of the electroweak phase transition. We also
discuss the range of validity of our approximations and compare this with other
approaches.
\end{abstract}
\vfill{
CERN-TH.6577/92
\newline
\noindent
revised - June 1993}
\end{titlepage}
\vskip2truecm

{\bf 1.}
In analogy with other symmetry-breaking phenomena in condensed matter
physics, the electroweak gauge symmetry $SU(2) \times U(1)$, which in
the Standard Model is spontaneously broken by the vacuum expectation
value $\phi$ of the Higgs field, is expected to be restored at sufficiently
high temperatures [\ref{kl1}]. Several years after the development of the
theoretical framework for a quantitative treatment of the subject [\ref{old}],
the interest in the electroweak phase transition has been revived by
the observation [\ref{sphal}] that the rate of anomalous $B$-violating
processes is unsuppressed at sufficiently high temperatures. This has
focused attention on the possibility that the cosmological baryon asymmetry
might be generated by physics at the Fermi scale (for recent reviews and
further references, see e.g. refs. [\ref{reviews}]).

One of the basic tools for the discussion of the electroweak phase
transition is $\veff$, the one-loop temperature-dependent effective
potential, improved by appropriate resummations of the most important
infrared-dominated higher-loop contributions, which jeopardize the
conventional perturbative expansion in the relevant range of temperature
and field values [\ref{old},\ref{books},\ref{fendley}]. A recent,
explicit computation of the Standard Model $T$-dependent one-loop potential
[\ref{ah}] was soon followed by an improved computation, which resummed the
leading, $\phi$-independent, $O(T^2)$ contributions to the effective masses
of scalar and vector bosons [\ref{carrington}]. Subsequent attempts to
include subleading contributions to the $T$-dependent effective
masses [\ref{shapo},\ref{brahm}] then originated a lively theoretical
debate [\ref{shapo}--\ref{buch}] about the nature of the electroweak phase
transition.

In this work we continue the discussion of subleading contributions to the
$T$-dependent effective masses of scalar and vector bosons in the Standard
Model. In particular, we compute the one-loop self-energies in the
zero-momentum limit, in the 't~Hooft-Landau gauge and in the approximation
$\lambda = g' = 0$. By solving the corresponding set of gap equations, and
keeping both leading and subleading terms in the high-temperature expansion of
the self-energies, we find that a non-vanishing magnetic mass is generated for
the $SU(2)$ gauge bosons, $m_T(0) = g^2 T /(3 \pi)$. We discuss its possible
implications for the nature of the electroweak phase transition. We also
comment on the range of validity of our calculational framework, and on the
possibility of consistently including subleading contributions in the
computation of $\veff$.

\vspace{7 mm}

{\bf 2.}
We begin by summarizing some well-known results, which will be useful for
the subsequent discussion. Here and in the following, we work in the
't~Hooft-Landau gauge. The spin-0 fields of the Standard Model are
described by the $SU(2)$ doublet
\be
\label{doublet}
\Phi = {1 \over \sqrt{2}} \left(
\begin{array}{c}
\chi_1 + i \chi_2
\\
\phi + h + i \chi_3
\end{array}
\right) \, ,
\ee
where $\phi$ is an arbitrary real constant background, $h$ is the Higgs field,
and $\chi_a$ ($a=1,2,3$) are the three Goldstone bosons. In terms of the
background field $\phi$, the tree-level potential is $V_{tree}(\phi) = - (\mu^2
/ 2) \phi^2 + (\lambda / 4) \phi^4$, with positive $\lambda$ and $\mu^2$, and
the tree-level minimum corresponds to $v^2 = \mu^2 / \lambda$. The spin-0
field-dependent masses are $m_h^2 (\phi) =  3 \lambda \phi^2 - \mu^2$ and
$m_{\chi}^2 (\phi) = \lambda \phi^2 - \mu^2$, so that $m_h^2 (v) = 2 \lambda
v^2 = 2 \mu^2$, $m^2_{\chi} (v) = 0$. The gauge bosons contributing to the
one-loop effective potential are $W^{\pm}$ and $Z$, with tree-level
field-dependent masses $m_W^2 (\phi) = (g^2 / 4) \phi^2$ and $m_Z^2 (\phi) =
[(g^2 + g'^2)/4] \phi^2$. Finally, the only fermion that can give a
significant contribution to the one-loop effective potential is the top quark,
with a field-dependent mass $m_t^2 (\phi) = (h_t^2 / 2) \phi^2$, where $h_t$ is
the top-quark Yukawa coupling.

The temperature-dependent one-loop effective potential can be calculated
according to standard techniques [\ref{kl1},\ref{old},\ref{books}].
Here and in the following, field-independent contributions will be
systematically neglected. The one-loop potential can be decomposed
into a zero-temperature and a finite-temperature part,
\be
\label{decomp}
V^{(1)} [m_i^2(\phi),T] =
V^{(1)} [m_i^2(\phi),0] +
\Delta V^{(1)} [m_i^2(\phi),T]
\, ,
\ee
which are finite functions of the renormalized fields and parameters,
once infinities are removed from the zero-temperature sector by means
of appropriate counterterms. Imposing renormalization conditions that
preserve the tree-level values of $v$ and $m_h(v)$, we can write
\be
\label{v0sm}
\begin{array}{rl}
V^{(1)} [m_i^2(\phi),0] & = {\displaystyle
\sum_{i=h,W,Z,t} {n_i \over 64 \pi^2} \left[
m_i^4 (\phi) \left( \log {m_i^2 (\phi) \over m_i^2 (v)}
- {3 \over 2} \right) + 2 m_i^2 (\phi) m_i^2 (v)  \right] }\\
& {\displaystyle + {n_{\chi} \over 64 \pi^2}
m_{\chi}^4 (\phi) \left[ \log {m_{\chi}^2(\phi) \over
m_h^2(v)} - {3 \over 2} \right] } \, ,
\end{array}
\ee
where $n_h=1$, $n_{\chi}=3$, $n_W=6$, $n_Z=3$, $n_t = - 12$, and we have used
the infinite running of the Higgs mass, from $p^2=0$ (where the effective
potential is defined) to $p^2=m_h^2$ (where the physical Higgs mass is
defined), in order to cancel the logarithmic infinity from the massless
Goldstone bosons at the minimum [\ref{ah},\ref{brahm}].

The finite-temperature part of the one-loop effective potential can be
written as
\be
\label{vtsm}
\Delta V^{(1)} [m_i^2(\phi),T] =
{T^4 \over 2 \pi^2} \left[
\sum_{i=h,\chi,W,Z} n_i \, J_+ (y_i^2) + n_t \, J_- (y_t^2)
\right] \, ,
\ee
where
\be
\label{ypsilon}
y^2 \equiv { m^2 (\phi) \over T^2 } \, ,
\;\;\;\;\;
J_{\pm} (y^2) \equiv \int_0^{\infty} dx \, x^2 \,
\log \left( 1 \mp e^{- \sqrt{x^2 + y^2}} \right) \, .
\ee
In view of the following discussion, we
write down the high-temperature expansion of eq.~(\ref{vtsm}),
\be
\label{vtsmht}
\begin{array}{c}
{\displaystyle
\Delta V^{(1)} [m_i^2(\phi),T] =
\sum_{i=h,\chi,W,Z} n_i \left\{
{m_i^2(\phi) T^2 \over 24}
-
{m_i^3(\phi) T \over 12 \pi}
-
{m_i^4(\phi) \over 64 \pi^2}
\left[ \log {m_i^2(\phi) \over T^2} - 5.4076 \right]
\right\}
}
\\
\\
{\displaystyle
\phantom{\Delta V^{(1)} [m_i^2(\phi),T] = \sum_{i=h,\chi,W,Z}}
-
n_t \left\{
{m_t^2(\phi) T^2 \over 48}
+
{m_t^4(\phi) \over 64 \pi^2}
\left[ \log {m_t^2(\phi) \over T^2} - 2.6350 \right] \right\} + \ldots \, ,
}
\end{array}
\ee
where the ellipsis stands for terms $O[m^6 (\phi)/T^2]$ or higher. Notice
that the $m^4 \log m^2$ terms cancel between the $T=0$ and the
$T \ne 0$ contributions in the high-temperature expansion. Notice also
that there is no $m_t^3(\phi)$ term in the high-temperature expansion
of the top-quark contribution to the one-loop potential, since there
cannot be fermionic modes of zero Matsubara frequency.

If one tries to use the full $T$-dependent one-loop effective potential, one
faces a number of difficulties [\ref{old},\ref{books},\ref{fendley}]: for
small values of $\phi$ [$\phi^2 < \mu^2 / \lambda$ for the Goldstone bosons,
$\phi^2 < \mu^2 /(3 \lambda)$ for the Higgs boson], the field-dependent
squared masses $m_h^2(\phi)$ and $m^2_{\chi}(\phi)$ become negative,
leading to a complex one-loop potential $V_{tree} + V^{(1)} [m_i^2(\phi),0]
+ \Delta V^{(1)} [m_i^2(\phi),T]$. More generally, for small values of the
field-dependent masses $m_i^2(\phi)$ ($i=h,\chi,W,Z$), the conventional loop
expansion is jeopardized by the (power-like) infrared behaviour of loop
diagrams at finite temperature. As will be recalled later, an appropriate
resummation of the leading infrared contributions from higher-loop diagrams
can remove this illness. Already at this level, however, we can take advantage
of the fact that for small values of the Higgs mass, $m_h^2 \ll m_W^2,
m_Z^2,m_t^2$,
the dominant contributions to the effective potential are those coming
from top quark and gauge boson loops. Since the field-dependent masses
$m_W^2(\phi)$, $m_Z^2(\phi)$ and $m_t^2(\phi)$ are positive for
any $\phi \ne 0$, we can consider, for the sake of discussion, a one-loop
finite-temperature potential in which only the $W$, $Z$ and top-quark loops
are taken into account. The behaviour of such a potential at the critical
temperature is described by the long-dashed line in fig.~1, for the
representative choice of parameters $m_h = 60 \gev $, $m_t = 150 \gev $.
We can see that for our parameter choice the naive one-loop potential
describes a first order phase transition, with the coexistence of two
degenerate minima occurring at $T \sim 95 \gev $.

\vspace{7 mm}

{\bf 3.}
We now wish to improve the one-loop result by including the most important
higher-loop contributions [\ref{old},\ref{books}]. As can be shown by simple
power-counting arguments, the higher-loop diagrams with the worst infrared
behaviour are the so-called daisy diagrams. They are generated
by iterated self-energy insertions, in the infrared limit $p^0 = 0$, $\vec{p}
\to 0$, on the one-loop diagrams carrying the modes of zero Matsubara
frequency. The $n=0$ contributions to the finite-temperature part of the
one-loop effective potential are nothing else than the $m^3 T$ terms in
the high-temperature expansion of eq.~(\ref{vtsmht}). Resumming the daisy
diagrams amounts then, as we shall see, to performing the
replacements $m_i^2(\phi) \to \mbar_i^2 (\phi, T)$ in the $m_i^3(\phi) T$
terms of eq.~(\ref{vtsmht}), where the effective masses $\mbar_i^2 (\phi, T)$
can be computed once the one-loop $T$-dependent self-energies are known.
We then begin by computing the $T$-dependent self-energies, for the different
bosonic fields that contribute to the one-loop effective potential, in the
infrared limit $p^0 = 0$, $\vec{p} \to 0$.
For simplicity, we compute them in the approximation $\lambda \to 0$ (i.e.
$m_h^2 \ll m_W^2,m_Z^2,m_t^2$) and $g' \to 0$, and we neglect their $T=0$
contributions. These approximations, which have been adopted in recent
analyses [\ref{shapo},\ref{brahm},\ref{bbh}], allow us to work with diagonal
one-loop self-energies, and are expected to reproduce the main
qualitative features of the result correctly.

The self-energies for the scalar fields $h$ and $\chi$, denoted by $\Pi_h$
and $\Pi_{\chi}$, are given by
\be
\label{polh}
\Pi_h = {3 g^2 \over 8 \pi^2} \left\{ T^2
\left[ I_+ (y_L^2) + 2 I_+ (y_T^2) \right] +
{g^2 \phi^2 \over 2} \left[
I_+' (y_L^2) + 2 I_+' (y_T^2) \right] \right\}
+
{3 h_t^2 T^2 \over \pi^2} \left[ I_- (y_t^2) +
{h_t^2 \phi^2 \over T^2} I_-' (y_t^2) \right] \, ,
\ee
and
\be
\label{polchi}
\Pi_{\chi} = {3 g^2 T^2 \over 8 \pi^2}
\left[ I_+ (y_L^2) + 2 I_+ (y_T^2) \right]
+
{3 h_t^2 T^2 \over \pi^2} I_- (y_t^2) \, ,
\ee
where $I_{\pm} ( y^2 ) = \pm 2 [d J_{\pm} (y^2) / d y^2]$, $I_{\pm}' ( y^2 ) =
d I_{\pm} / d y^2$.

In the chosen approximation, the $W$ and the $Z$ are degenerate in mass
\footnote{To compensate for this fact, we shall adopt in the following the
numerical value $g^2 = 4 \frac{2 m_W^2 + m_Z^2}{3 v^2}$.}, and their propagator
can be decomposed into a transverse and a longitudinal part as
\be
\label{vprop}
D_{\mu \nu} (p) =
{ T_{\mu \nu} \over {p^2 - m_T^2}}
+
{ L_{\mu \nu} \over {p^2 - m_L^2}} \, ,
\ee
where ($i,j=1,2,3$)
\be
\label{tlprop}
T_{\mu \nu} = - g_{\mu i} \left( g^{ij} + {p^i p^j \over \vec{p}^{\; 2}}
\right) g_{j \nu} \, ,
\;\;\;\;\;
L_{\mu \nu} = {p_{\mu} p_{\nu} \over p^2 }
- g_{\mu \nu} - T_{\mu \nu} \, ,
\ee
satisfy the orthogonality conditions $T_{\mu \nu} T^\nu_{\lambda} = - T_{\mu
\lambda}$, $L_{\mu \nu} L^\nu_{\lambda} = - L_{\mu \lambda}$, $T_{\mu \nu}
L^\nu_{\lambda} = 0$. The transverse and longitudinal masses are equal at the
tree level, $m_T^2 (\phi)=m_L^2(\phi)=g^2 \phi^2 / 4 \equiv m^2 (\phi)$.
However, they will get different one-loop contributions at finite temperature,
since for $T \ne 0$ the longitudinal and transverse components act in practice
as independent degrees of freedom in Feynman diagrams. We will then
decompose the gauge boson self-energy, according to (\ref{vprop}), as
[\ref{books}] $\Pi_{\mu \nu} = L_{\mu \nu} \Pi_L + T_{\mu \nu} \Pi_T$,
where, in the infrared limit, $\Pi_L(0) = \Pi^0_0 (0)$, $\Pi_T (0) = (1/3)
\sum_i \Pi_i^i (0)$.

For the longitudinal polarization, we obtain
\be
\label{pil}
\begin{array}{ll}
\Pi_L = &
{g^2 T^2 \over \pi^2} \left\{
2 \left[ I_+ (y_L^2) + 2 I_+ (y_T^2) \right]
-
2 \left[ y_L^2 I_+' (y_L^2) +
2 y_T^2 I_+' (y_T^2) \right]
-
{3 \over y_L^2} \left[ J_+ (y_L^2) - J_+ (0) \right]
\right.
\\ & \\ &
\left.
- \; {1 \over 2} I_+ (0)
+ { 5 I_+ (y_{\chi}^2) +  I_+ (y_h^2) \over 8 }
- {y_{\chi}^2 \over 2} I_+' (y_{\chi}^2)
+ {3 \over 2} {J_+ (y_h^2) - J_+ (y_{\chi}^2) \over
y_h^2 - y_{\chi}^2}
-
{1 \over 2} { y_h^2 I_+ (y_h^2) - y_{\chi}^2 I_+ (y_{\chi}^2) \over
y_h^2 - y_{\chi}^2} \right\}
\\ & \\ &
+ \; {3 g^4 \phi^2 \over 8 \pi^2 (y_L^2 - y_h^2)}
\left[ { J_+ (y_L^2) - J_+ (0) \over y_L^2}
- { J_+ (y_h^2) - J_+ (0) \over y_h^2} \right]
+
g^2 T^2 \left\{
{7 \over 8} + {3 \over 2 \pi^2}
\left[  I_- (y_t^2) - {y_t^2 \over 2} I_-' (y_t^2) \right] \right\} \, ,
\end{array}
\ee
and for the transverse polarization
\be
\label{pit}
\begin{array}{ll}
\Pi_T = &
{g^2  T^2 \over \pi^2} \left\{
- {1 \over 3}  I_+ (y_L^2) - \frac{2}{3} I_+ (y_T^2) + {1 \over 2} I_+ (0) +
{J_+ (y_L^2) - J_+ (0) \over y_L^2}
+
{I_+ (y_h^2) +  I_+ (y_{\chi}^2) \over 8}
-
{J_+ (y_h^2) - J_+ (y_{\chi}^2) \over
2 (y_h^2 - y_{\chi}^2)}
\right.
\\ & \\ &
+
\left.
{3 y_t^2 \over 4} I_-' (y_t^2) \right\}
+
{g^4 \phi^2 \over 8 \pi^2 (y_T^2 - y_h^2)}
\left[ I_+ (y_T^2) - I_+ (y_h^2)
- { J_+ (y_T^2) - J_+ (0) \over y_T^2}
+ { J_+ (y_h^2) - J_+ (0) \over y_h^2}
\right] \, .
\end{array}
\ee
The high-temperature expansion of the self-energies in eqs.~(\ref{polh}),
(\ref{polchi}), (\ref{pil}), (\ref{pit}) reads
\be
\label{verycl1}
\Pi_h [m_i^2(\phi),T] =
\left( {3 \over 16} g^2 + {1 \over 4} h_t^2 \right) T^2
- {3 g^2 \over 16 \pi } \left( m_L + 2 m_T \right) T
- {3 g^4 \phi^2 \over 32 \pi }
\left( {1 \over m_T} + {1 \over 2 m_L} \right) T
+ \ldots \, ,
\ee
\be
\label{verycl2}
\Pi_{\chi} [m_i^2(\phi),T] =
\left( {3 \over 16} g^2 + {1 \over 4} h_t^2 \right) T^2
- {3 g^2 \over 16 \pi } \left( m_L + 2 m_T \right) T
+ \ldots \, ,
\ee
\be
\label{verycl3}
\Pi_L [m_i^2(\phi),T] = {11 \over 6} g^2 T^2
- {g^2 \over 16 \pi } \left( m_h + 3 m_{\chi} + 16 m_T \right) T
- {g^4 \phi^2 \over 16 \pi } {T \over m_L + m_h} +  \ldots \, ,
\ee
\be
\label{verycl4}
\Pi_T [m_i^2(\phi),T] =
{g^2 \over 3 \pi} m_T T
+ {g^2 \over 12 \pi } \left( {m_h +  m_{\chi} \over 4}
- {m_h m_{\chi} \over m_h + m_{\chi}} \right) T
- {g^4 \phi^2 \over 24 \pi } {T \over m_T + m_h} +  \ldots \, .
\ee

Using the one-loop self-energies given above, one can improve the one-loop
result of eq.~(\ref{decomp}) by considering the daisy vacuum diagrams. Using
the previous decompositions of the propagator and of the self-energy, we can
write
\be
\label{gaupro}
D_{\mu \nu} \Pi_{\lambda}^{\nu}
=
- L_{\mu \lambda} {\Pi_L \over p^2 - m_L^2}
- T_{\mu \lambda} {\Pi_T \over p^2 - m_T^2} \, .
\ee
The contribution to the effective potential of the daisy diagrams obtained from
gauge-boson loops is then
\be
\label{vecdaisy}
- {T \over 2} \int
{d^3 \vec{p} \over {(2 \pi)^3}}
{\rm Tr} \sum_{N=1}^{\infty}
{1 \over N}
\left[
L_{\mu \nu} {\Pi_L \over  \vec{p}^{\; 2} + m^2_L}
+
T_{\mu \nu} {\Pi_T \over  \vec{p}^{\; 2} + m^2_T}
\right]^N
 \, ,
\ee
where the trace is over Lorentz indices (in the approximation $g'=0$,
the one-loop self-energy matrix of neutral gauge bosons is diagonal
whereas, for $g' \ne 0$, mixing effects should be properly taken into account).
Using again the orthogonality properties of $L_{\mu \nu}$ and $T_{\mu \nu}$,
the final result is
\be
\label{vzwdai}
\Delta V^{daisy} =
- {T \over 2}
\sum_{i=h,\chi,L,T} n_i
\int {d^3 \vec{p} \over {(2 \pi)^3}}
\sum_{N=1}^{\infty} {1 \over N}
\left[ - {\Pi_i \over  \vec{p}^{\; 2} + m^2_i}
\right]^N \, .
\ee
Summarizing, the effective potential improved by the daisy diagrams
is\footnote{As for the top quark contribution to $\veff$, we did not
perform any resummation, since fermion loops do not suffer from the
infrared problems associated with the zero Matsubara frequencies.}
\be
\label{veffdai}
V_{eff} = V_{tree} + V^{(1)} [ m_i^2 (\phi) , T ]
-
{T \over 12 \pi} \sum_{i=h,\chi,L,T}
n_i \left[ \mbar_i^3 (\phi,T) - m_i^3 (\phi) \right] \, ,
\ee
where
\be
\label{gapdaism}
\mbar_i^2 (\phi,T) = m_i^2 (\phi)  + \Pi_i[ m_j^2(\phi) , T ] \, .
\ee
For consistency, in evaluating [\ref{carrington},\ref{dine},\ref{arnold}] the
daisy-improved effective potential of eq.~(\ref{veffdai}), one has to keep only
the $T^2$ terms in the self-energies of eqs.~(\ref{verycl1})--(\ref{verycl4}).
This amounts to using the following set of $T$-dependent masses
\be
\label{carsm}
\mbar_h^2 - m_h^2 = \mbar_{\chi}^2 - m_{\chi}^2 = {1 \over 4}
\left( {3 \over 4} g^2 + h_t^2 \right) T^2 \, ,
\;\;\;\;\;
\mbar_L^2 = m^2 + {11 \over 6} g^2 T^2 \, ,
\;\;\;\;\;
\mbar_T^2 = m^2 \, .
\ee
The resulting effective potential at the critical temperature is displayed as
the short-dashed line in fig.~1, for the same parameter choices as before. One
can notice that the effective potential corresponding to eqs.~(\ref{veffdai})
and (\ref{carsm}) still describes a first-order phase transition. However, one
can also observe [\ref{dine}] the characteristic reduction by a factor of
$\sim 2/3$ in the non-trivial vacuum expectation value of the field $\phi$,
at the temperature $T_c \sim 97 \gev$ at which the two minima become
degenerate. This point can be qualitatively understood by looking at the
coefficient of the $\phi^3$ term in the high-temperature expansion of the
effective potential.

\vspace{7 mm}

{\bf 4.}
To improve over the previous approximation, one could try to
keep also the subleading terms in the high-temperature expansion of the
self-energies in eqs.~(\ref{verycl1})--(\ref{verycl4}),
obtaining the effective masses
\be
\label{shasm}
\begin{array}{rcl}
\mbar_h^2 & = &  m_h^2 +
{1 \over 4} \left( {3 \over 4} g^2 + h_t^2 \right) T^2
- {9 \over 16 \pi} g^3 \phi T \, ,
\\
& & \\
\mbar_{\chi}^2 & = & m_{\chi}^2 +
{1 \over 4} \left( {3 \over 4} g^2 + h_t^2 \right) T^2
- {9 \over 32 \pi} g^3 \phi T \, ,
\\
& & \\
\mbar_L^2 & = &  m^2 +
{11 \over 6} g^2 T^2 - {5 \over 8 \pi} g^3 \phi T \, ,
\\
& & \\
\mbar_T^2 & = &
m^2 + {1 \over 12 \pi} g^3 \phi T \, .
\end{array}
\ee
Doing so, one would obtain linear terms in $\phi$ from the effective
potential of eq.~(\ref{veffdai}). From the gauge boson sector,
one would get $\sqrt{33 / 2} [5 / (64 \pi^2)] g^4  \phi T^3$,
a positive linear term near the origin, as observed in ref.~[\ref{shapo}].
Including the Higgs and Goldstone boson sectors, one would obtain the
additional linear term $[45 / (512 \pi^2)] g^3 \sqrt{ (3/4) g^2 + h_t^2} \phi
T^3$. Both results quoted above originate from subleading
contributions to the self-energies. They are a consequence of the set of
(daisy) diagrams considered to improve the one-loop result: as shown
in ref.~[\ref{eqz}], the daisy resummation is an inconsistent procedure
at subleading order.

Na\"{\i}ve power-counting arguments would suggest that, to include subleading
corrections to the effective potential, one has to consider the full set of
super-daisy diagrams, where all bubbles belonging to daisy diagrams have
bubble-corrected propagators. This amounts to solving a system of gap
equations, of the form
\be
\label{gapsm}
\ov{m}^2_i =
m_i^2 + \Pi_i (\mbar_h , \mbar_{\chi} , \mbar_L , \mbar_T ) \, ,
\;\;\;\;\;
(i = h, \chi, L, T) \, ,
\ee
and to writing down for the effective potential the formal expression
(\ref{veffdai}), but now with the masses $\mbar_i^2$
given by the solution\footnote{When solving the gap equations (\ref{gapsm}),
the results obtained by using our approximate expressions for the
self-energies, eqs.~(\ref{verycl1})--(\ref{verycl4}), disagree with those of
ref.~[\ref{bbh}]. We can identify two sources of disagreement.
First, ref.~[\ref{bbh}] does not distinguish between transverse and
longitudinal vector boson masses on the right-hand side of eq.~(\ref{gapsm}).
Second, when solving the gap equations for $\mbar_L$ and $\mbar_T$,
ref.~[\ref{bbh}] neglects the $g^2 T^2$ and $h_t^2 T^2$ corrections to the
Higgs and Goldstone boson masses, which make them different from zero even
in the limit $\lambda \to 0$. At subleading order this is not correct,
since the masses appearing in the self-energies of eq.~(\ref{gapsm}) are
the unknown solutions of the gap equations.} of eq.~(\ref{gapsm}).

In the improved theory, the expansion of the effective potential is controlled
by the set of parameters
\be
\label{abt}
\ov{\alpha}_i = {g^2 \over 2 \pi} {T^2 \over \mbar_i^2} \, ,
\;\;\;\;\;
\bbar_i = {g^2 \over 2 \pi} {T \over \mbar_i} \, ,
\;\;\;\;\;
\gamma = {\phi^2 \over T^2} \, ,
\ee
where $i=h,\chi,L,T$ and $\mbar_i^2$ are the improved masses that solve
the gap equations (\ref{gapsm}).  Accordingly, we
can write the solution to the gap equations (\ref{gapsm}) to leading order
[$O(\bar{\alpha}_i)$ for $i=h,\chi,L$, $O(\bbar, \, \ov{\alpha}_T \bbar
\gamma)$ for $\mbar_T$]  as
\be
\label{oursol}
\begin{array}{rcl}
\mbar_h^2 & = &  m_h^2 (\phi) +
\left( {3 \over 16} g^2 + {1 \over 4} h_t^2 \right) T^2 \, ,
\\
& & \\
\mbar_{\chi}^2 & = & m_{\chi}^2 (\phi) +
\left( {3 \over 16} g^2 + {1 \over 4} h_t^2 \right) T^2 \, ,
\\
& & \\
\mbar_L^2 & = &  m^2 (\phi) + {11 \over 6} g^2 T^2 \, ,
\\
\end{array}
\ee
and
\be
\label{mtfirst}
\mbar_T^2 = {\displaystyle  m^2(\phi)+ \frac{g^2T}{3\pi} m(\phi)+
\frac{g^2T}{48\pi} \frac{(\mbar_h-\mbar_{\chi})^2}{\mbar_h+\mbar_{\chi}}
-\frac{g^4 \phi^2}{24\pi} \frac{T}{m(\phi)+\mbar_h} } \, .
\ee
Notice that the solution (\ref{mtfirst}) for $\mbar_T$ improves over
the tree-level solution $g \phi / 2$ by the inclusion of $O(\bbar, \,
\ov{\alpha}_T \bbar \gamma)$ terms. The latter provide the leading plasma
contributions to $\mbar_T$ and should therefore be kept.

We can see from eqs.~(\ref{verycl1}) and (\ref{verycl2}) that
at the origin, $\phi=0$,
$\Pi_h(m_i^2(0),T)=\Pi_{\chi}(m_i^2(0),T)$ and therefore, since
$m_h^2(0)=m_{\chi}^2(0)$, the general solutions to the gap
equations (\ref{gapsm}) satisfy the condition
$\mbar_h(0)=\mbar_{\chi}(0)$. Using now, from eq.~(\ref{verycl4}), the
fact that $\Pi_T(\mbar_i^2(0),T)=\frac{g^2}{3\pi} \mbar_T T$, the gap
equation for $\mbar_T$ at $\phi=0$ can be simply written as
\begin{equation}
\label{eq0}
\mbar_T^2=\frac{g^2T}{3\pi} \mbar_T \ ,
\end{equation}
with the two solutions
\be
\label{sol1}
\mbar_T(0)=0
\ee
and
\be
\label{sol2}
\mbar_T(0)=\frac{g^2T}{3\pi} \, .
\ee
Solution (\ref{sol1}) coincides with (\ref{mtfirst}) at $\phi=0$.
It is reached when solving the gap equation for $\mbar_T$
\be
\label{impeq}
\begin{array}{rcl}
\mbar_T^2 & = & {\displaystyle m^2(\phi)+ \frac{g^2T}{3\pi} \mbar_T+
  \frac{g^2T}{48\pi}
\frac{(\mbar_h-\mbar_{\chi})^2}{\mbar_h+\mbar_{\chi}}
-\frac{g^4 \phi^2}{24\pi} \frac{T}{\mbar_T+\mbar_h} }
\end{array}
\ee
to {\it first} order in $\ov{\beta}, \, \ov{\alpha}_T \bbar \gamma$.
In that case the validity of the perturbative expansion in the improved
theory breaks down near the origin, since $\overline{\beta}_T \rightarrow
\infty$ when $\phi \rightarrow 0$. However, by going to all orders in
$\overline{\beta}$ one can reach the (non-perturbative) solution
(\ref{sol2}) and avoid the singularity of $\overline{\beta}_T$ (make
the improved perturbative expansion more reliable) at the origin.

The solution to the gap equation (\ref{impeq}), which is
continuously connected to (\ref{sol2}), is given by
\be
\label{impsol}
\mbar_T  =  2 \sqrt{Q(\phi)} \cos \left[  {1 \over 3} \theta (\phi)
\right] + {1 \over 3} \left( {g^2 T \over 3 \pi} - \mbar_h \right) \, ,
\ee
where the functions $Q(\phi)$ and $\theta(\phi)$ are defined as
\be
\label{qfun}
Q (\phi) = {1 \over 9} (a^2 + 3 b) \,
\;\;\;\;\;
\cos \theta (\phi) = {1 \over 2} { 9 a b + 2 a^3 - 27 c \over
(a^2 + 3 b)^{3/2}} \, ,
\ee
with
\be
\label{apar}
a = {g^2 T \over 3 \pi} + 2 \mbar_h \, ,
\;\;\;\;\;
c = {g^4 \over 24 \pi} \phi^2 T \, ,
\ee
\be
\label{bpar}
b = m^2 (\phi) + {g^2 T \over 48 \pi}
{(\mbar_h - \mbar_{\chi})^2 \over \mbar_h + \mbar_{\chi}}
- \left( {g^2 T \over 3 \pi} + \mbar_h \right) \mbar_h \, .
\ee
For temperatures close to the critical temperature, $T \sim 100 \gev$,
our solution would give $\mbar_T (0) \sim 5 \gev$, with $\mbar_T^2 (\phi)$
quickly approaching $m^2(\phi)$ for increasing values of $\phi$. It is also
interesting to consider the parameters $\bbar_i (\phi)$ $(i=h,\chi,L,T)$, which
can help to identify the region where our approximations are reliable. Whilst
$\bbar_{h,\chi,L} \ll 1$ for all values of $\phi$, near $\phi=0$ it is $\bbar_T
\sim 1.5$, so that our improved perturbative expansion is no longer reliable,
and unaccounted for non-perturbative effects could significantly change our
results. However, for increasing $\phi$, also $\bbar_T$ moves fast into the
perturbative regime.

Considering $O(\bbar, \, \ov{\alpha}_i \bbar \gamma)$, subleading
corrections to $\mbar_h$, $\mbar_{\chi}$, $\mbar_L$ would amount to adding
\be
\label{adding}
\begin{array}{rcl}
\Delta \mbar_h^2 & = &  {\displaystyle
- {3 g^2 \over 16 \pi} \left( \mbar_L + 2 \mbar_T \right) T
- {3 g^4 \phi^2  \over 32 \pi} \left( {1 \over \mbar_T} + {1 \over 2 \mbar_L}
\right) T } \, ,
\\
& & \\
\Delta \mbar_{\chi}^2 & = & {\displaystyle
- {3 g^2 \over 16 \pi} \left( \mbar_L + 2 \mbar_T \right) T } \, ,
\\
& & \\
\Delta \mbar_L^2 & = &  {\displaystyle
- {g^2 \over 16 \pi} \left( \mbar_h + 3 \mbar_{\chi} + 16 \mbar_T \right) T
- {g^4 \phi^2  \over 16 \pi} {T \over \mbar_L + \mbar_h} } \, ,
\end{array}
\ee
where $\mbar_i$ ($i=h,\chi,L,T$) are the solutions in
eqs.~(\ref{oursol}) and (\ref{impsol})--(\ref{bpar}).
Notice that the total squared masses $\mbar_i^2 + \Delta \mbar_i^2$
($i=h,\chi,L,T$; $\Delta \mbar_T^2=0$), as given by eqs.~(\ref{oursol}),
and (\ref{impsol})--(\ref{adding}), do not have any linear term in $\phi$,
unlike (23), and so would not give rise to any linear term in the effective
potential, as anticipated.

Before proceeding, some comments on the previous results are in order. The
generation of a non-vanishing transverse mass at $\phi=0$, eq.~(\ref{sol2}),
does not contradict any general result. It is known [\ref{klimov}] that in
scalar electrodynamics at finite temperature the transverse gauge-boson
mass is equal to zero. However, this theorem does not apply to non-Abelian
gauge theories, and it is easy to check that the transverse mass of
eq.~(\ref{sol2}) is originated by the quartic non-Abelian gauge coupling.
We should also stress that in our calculational framework the masses $\mbar_i$
are gauge-dependent at subleading order and do not have a direct physical
meaning, but are just an intermediate result in the computation of the improved
effective potential. Calculations similar to ours (even if performed in
different gauges) found a non-vanishing transverse mass in finite-temperature
QCD [\ref{kal}]. After the appearance of the first version of the present
paper, our result was confirmed [\ref{buchbis}] by an independent calculation
along the same lines\footnote{Ref.~[\ref{buchbis}] also pointed out a subtlety
in the decomposition of the gauge boson self-energy, leading to a
correction factor $(2/3)$, which we have now included.}. Other techniques for
estimating the magnetic mass $m_{mag}$ have been tried in the literature, with
different results. Lattice calculations [\ref{latticebis}] give estimates of
$m_{mag}$ that [rescaled to the $SU(2)$ case when appropriate] range from $\sim
0.25 g^2 T$ to $\sim 3 g^2 T$. Resummation techniques using a
gauge-invariant propagator [\ref{corn}] find an approximate lower bound
$m_{mag} \simgt 0.6 g^2 T$. Exact zero-momentum sum rules in $d=3$ gauge
theory lead to the estimate [\ref{cornbis}] $m_{mag} \sim 0.23 g^2 T$. A
recent estimate of $m_{mag}$ based on a semiclassical method [\ref{magnetic}]
also gives a very similar result.

\vspace{7 mm}

{\bf 5.}
We now want to discuss the possibility of improving the effective potential by
making use of the previous results. As we already mentioned, an attempt to
include subleading corrections coming from higher-loop diagrams is the
so-called superdaisy approximation: first one solves the gap equation in the
infrared limit at subleading order, then one plugs the solution $\mbar_i$ into
the expression (\ref{veffdai}). This procedure, however, has a number of
shortcomings, which we are now going to describe.

The first one is [\ref{books},\ref{eqz}] that the combinatorics of superdaisy
diagrams do not fit the shifting in the $m^3$ term, eq.~(\ref{veffdai}), and
have to be compensated by a correction $\Delta V^{comb}$ to the effective
potential\footnote{In spite of recent claims [\ref{boydbis}], this
combinatorial mismatch is exactly the same whatever construction is used for
the effective potential, either the one based on vacuum diagrams or the one
based on the integration of the tadpole diagrams [\ref{eqz},\ref{quiros}].}.
If one works with the leading-order expressions (\ref{carsm}) for the
effective masses, the correct combinatorics is automatically reproduced
by the counting of different two-loop contributions corresponding to
just one propagator with zero Matsubara frequency, so $\Delta V^{comb} = 0$.
To subleading order, however, one should also include the two-loop diagrams
where all propagators correspond to zero Matsubara frequency: in this case
$\Delta V^{comb} \ne 0 $. For example, in the scalar theory discussed in
[\ref{eqz}] one would find $\Delta V^{comb} = (-7 \lambda)/(64 \pi^2) \cdot
m^2(\phi) T^2$, in full agreement with the recent results of ref.~[\ref{pi}],
once the effects of the cubic coupling are taken into account. In principle,
one can compute $\Delta V^{comb}$ at arbitrary order in the expansion parameter
$\bbar$, as explicitly shown in refs.~[\ref{pi},\ref{quiros}]. However, the
superdaisy approximation is certainly not accurate beyond subleading order
$\bbar$ [\ref{eqz}], so higher-loop contributions to $\Delta V^{comb}$ are only
of academical interest.

The second, more serious shortcoming of the superdaisy approximation, which was
recently stressed in ref.~[\ref{ae}], is related to the fact that the masses
obtained from the gap equation (\ref{gapsm}) are calculated in the
zero-momentum limit. In fact, diagrams with overlapping momenta give
logarithmic contributions to the effective potential that are missed by the
superdaisy approximation, but are also corrections belonging to subleading
order $\bbar$. In spite of some interesting recent attempts [\ref{amelino}] to
achieve consistency of the superdaisy approximation at subleading order, this
still remains, in our opinion, an open problem. In ref.~[\ref{ae}] a hybrid
method was proposed, where resummation limited to leading order is combined
with a full two-loop computation. We shall take here an alternative approach,
following the lines of ref.~[\ref{buchbis}]. Given the uncertainties in the
determination of the magnetic mass, and the absence of a consistent computation
of $\veff$ at subleading order $\bbar$, we shall attempt a phenomenological
parametrization of subleading effects by introducing a constant magnetic mass
$\mbar_T(0) = \gamma \cdot (g^2 T)/(3 \pi)$, and by replacing $m_T^2(\phi)$
with $\mbar^2_T(\phi) = \mbar_T^2(0) + m^2(\phi)$ in eq.~(\ref{veffdai}).

We plot as solid lines in fig.~1 the improved potentials corresponding
to our approach, at the critical temperature $T_c \sim 97 \gev$, for the same
choice of  $m_h$ and $m_t$ as before, and for some representative values of
$\gamma$. The result still shows, for $\gamma \simlt 2$, a first-order phase
transition, but weaker than that in the approximation corresponding to the
short-dashed line. The reason is that no temperature screening was considered
in the latter for $\mbar_T$. The subleading-order
screening parametrized by our approach translates into a further weakening of
the first-order phase transition, which depends on $m_h$ and $m_t$. We then
compare the different approximations for various values of the Higgs mass
$m_h$. We plot in fig.~2 the ratio $v(T_c)/T_c$ as a function of $m_h$;
the different curves have the same meaning as in fig.~1. We can see that, for a
fixed value of $\gamma$, the phase transition becomes second order [i.e.
$v(T_c)/T_c = 0$] for Higgs masses greater than a critical value. Also, for a
fixed value of $m_h$, there is a critical value of $\gamma$ beyond which the
phase transition becomes second order. For instance, for $m_h= 60 \gev$,
$m_t=150 \gev$ and $\gamma=3$, the phase transition is second order, as can be
seen from fig.~1. There is also a mild dependence of $v(T_c)/T_c$ on $m_t$,
which was not explicitly exhibited in the figures. For $m_h=60 \gev$ and
$m_t=130 \gev$ ($m_t=170 \gev$) the value of $v(T_c)/T_c$ increases (decreases)
by $\sim 20 \%$ with respect to its value at $m_t=150 \gev$. Anyhow, the main
result that can be read off fig.~2, i.e. the turnover to a second-order
phase transition, has to be taken {\it cum grano salis}, since it occurs at
large values of $m_h$, where our approximation is less reliable, or large
values of $\gamma$, which could be provided by non-perturbative effects not
accounted for by our calculation. On the other hand, we do not believe that
the inclusion of terms of order $\lambda$ and/or $g'$ in the self-energies
will qualitatively change the above picture.

\vspace{7 mm}

{\bf 6.}
In conclusion, we have analysed the nature of the phase transition and
the structure of the finite-temperature effective potential of the Standard
Model, $\veff$, including some higher-order infrared-dominated contributions.
We have improved over the existing computations by including subleading
contributions to the $T$-dependent effective masses of the transverse
polarizations of vector bosons, in the zero-momentum approximation.
In particular, we have found that in the 't Hooft-Landau gauge a non-vanishing
magnetic mass $\mbar_T(0) = g^2 T/(3 \pi)$ is generated. In the absence of a
consistent resummation procedure for the subleading higher-loop contributions
to $\veff$, we have studied the effects of a non-vanishing magnetic mass,
phenomenologically parametrized in terms of a fudge factor $\gamma$. We have
proved that, in our approach, subleading corrections {\em do not} lead to
linear $\phi$-terms in $\veff$. We have also found a tendency towards a
further weakening of the phase transition, with the possibility of a
second-order phase transition for sufficiently large values of $m_h$ and
$\gamma$. The computation of other subleading effects, associated with
overlapping momenta in two-loop graphs, gives however corrections going in the
opposite direction [\ref{ae},\ref{bd}]. Both these results should then be taken
as indications of the possible size of subleading effects, but none of the two
is fully consistent at subleading order. In any case, these corrections do not
seem to be sizeable enough to rescue the Standard Model as a candidate for
electroweak baryogenesis.

It would be good to have available more reliable calculational methods to deal
with non-perturbative effects. Some possibilities currently under investigation
are lattice computations at finite temperature [\ref{lattice}], the
$(1/N)$-expansion [\ref{oneovern}], the $\epsilon$-expansion [\ref{epsilon}]
and various $d=3$ effective theory methods [\ref{threed}], in particular the
average potential method [\ref{wetterich}], but none of them has yet given
conclusive results for the Standard Model case and the presently allowed values
of $m_t$ and $m_h$.

As a final remark, we would like to recall that the effective potential
describes the static properties of the phase transition, and that many other
subtle issues have to be addressed when discussing its dynamical aspects.

\section*{Acknowledgements}
We would like to thank  C.G.~Boyd, D.E.~Brahm, A.~Brignole, W. Buchm\"uller,
J.M.~Cornwall, K.~Farakos, V.~Jain, C.~Kounnas and M.E.~Shaposhnikov for useful
discussions and suggestions.

\newpage
\section*{References}
\begin{enumerate}
\item
\label{kl1}
D.A.~Kirzhnits and A.D.~Linde, Phys. Lett. 72B (1972) 471.
\item
\label{old}
S. Weinberg, Phys. Rev. D9 (1974) 3357;
L. Dolan and R. Jackiw, Phys. Rev. D9 (1974) 3320;
D.A.~Kirzhnits and A.D.~Linde, Sov. Phys. JETP 40 (1975) 628 and
Ann. Phys. 101 (1976) 195.
\item
\label{sphal}
F.R.~Klinkhamer and N.S.~Manton, Phys. Rev. D30 (1984) 2212;
V.A.~Kuzmin, V.A.~Rubakov and M.E.~Shaposhnikov, Phys. Lett. B155 (1985) 36.
\item
\label{reviews}
A.D.~Dolgov, Phys. Rep. 222 (1992) 309;
M.E.~Shaposhnikov, in {\em `Proceedings of the 1991 Summer School in High
Energy Physics and Cosmology'}, Trieste, 17 June--9 August 1991,
E.~Gava et al., eds. (World Scientific, Singapore, 1992), Vol.~1,
p.~338; A.G.~Cohen, D.B.~Kaplan and A.E.~Nelson, San Diego preprint
UCSD-PTH-93-02, BUHEP-93-4.
\item
\label{books}
D.J.~Gross, R.D.~Pisarski and L.G.~Yaffe,  Rev. Mod. Phys. 53 (1981) 43;
J.I.~Kapusta, {\em Finite-temperature Field Theory} (Cambridge University
Press, 1989);
A.D. Linde, {\em Particle Physics and Inflationary Cosmology} (Harwood,
New York, 1990).
\item
\label{fendley}
P.~Fendley, Phys. Lett. B196 (1987) 175.
\item
\label{ah}
G.W.~Anderson and L.J.~Hall, Phys. Rev. D45 (1992) 2685.
\item
\label{carrington}
M.E. Carrington, Phys. Rev. D45 (1992) 2933.
\item
\label{shapo}
M.E.~Shaposhnikov, Phys. Lett. B277 (1992) 324 and (E) B282 (1992) 483.
\item
\label{brahm}
D.E.~Brahm and S.D.H.~Hsu, Caltech preprints CALT-68-1762,
HUTP-91-A064 and CALT-68-1705, HUTP-91-A063.
\item
\label{dine}
M.~Dine, R.G.~Leigh, P.~Huet, A.~Linde and D.~Linde, Phys. Lett. B283
(1992) 319 and Phys. Rev. D46 (1992) 550.
\item
\label{arnold}
P.~Arnold, Phys. Rev. D46 (1992) 2628.
\item
\label{eqz}
J.R.~Espinosa, M.~Quir\'os and F.~Zwirner, Phys. Lett. B291
(1992) 115.
\item
\label{bbh}
C.G.~Boyd, D.E.~Brahm and S.D.~Hsu, Caltech preprint CALT-68-1795,
HUTP-92-A027, EFI-92-22.
\item
\label{buch}
W.~Buchm\"uller and T.~Helbig, Int. J. Mod. Phys. C3 (1992) 799;
W.~Buchm\"uller, T.~Helbig and D.~Walliser, preprint DESY 92-151.
\item
\label{klimov}
O.K.~Kalashnikov and V.V.~Klimov, Phys. Lett. 95B (1980) 423.
\item
\label{kal}
O.K.~Kalashnikov, Phys. Lett. B279 (1992) 367.
\item
\label{buchbis}
W.~Buchm\"uller, Z.~Fodor, T.~Helbig and D.~Walliser, preprint DESY 93-021.
\item
\label{latticebis}
A.~Billoire, G.~Lazarides and Q.~Shafi, Phys. Lett. B103 (1981) 450;
T.~De~Grand and D.~Toussaint, Phys. Rev. D25 (1982) 526;
G.~Lazarides and S.~Sarantakos, Phys. Rev. D31 (1985) 389;
G.~Koutsoumbas, K.~Farakos and S.~Sarantakos, Phys. Lett. B189 (1986) 173;
A.~Irb\"ack and C.~Peterson, Phys. Lett. B174 (1986) 99;
J.~Ambjorn, K.~Farakos and M.~Shaposhnikov, preprint CERN-TH.6508/92;
J.E.~Mandula and M.~Ogilvie, Phys. Lett. B201 (1988) 117;
M.~Teper, Phys. Lett. B289 (1992) 115;
C.~Bernard, Phys. Lett. B108 (1982) 431;
J.E.~Mandula and M.~Ogilvie, Phys. Lett. B185 (1987) 127.
\item
\label{corn}
J.M.~Cornwall, W.-S. Hou and J.E.~King, Phys. Lett. B153 (1985) 173.
\item
\label{cornbis}
J.M.~Cornwall, preprint UCLA/92/TEP/51.
\item
\label{magnetic}
T.S.~Bir\'o and B.~M\"uller, Duke University preprint DUKE-TH-92-42.
\item
\label{boydbis}
C.G.~Boyd, D.E.~Brahm and S.D.~Hsu, Caltech preprint CALT-68-1858,
HUTP-93-A011, EFI-93-22.
\item
\label{quiros}
M.~Quir\'os, Madrid preprint IEM-FT-71/93.
\item
\label{pi}
G.~Amelino-Camelia and S.-Y.~Pi, Phys. Rev. D47 (1993) 2356.
\item
\label{ae}
P.~Arnold and O.~Espinosa, Phys. Rev. D47 (1993) 3546.
\item
\label{amelino}
G.~Amelino-Camelia, Boston University preprint BUHEP-93-12.
\item
\label{bd}
J.E.~Bagnasco and M.~Dine, Santa Cruz preprint SCIPP-92-43.
\item
\label{lattice}
B.~Bunk, E.M.~Ilgenfritz, J.~Kripfganz and A.~Schiller, Phys. Lett.
B284 (1992) 371 and Bielefeld preprint BI-TP-92-46; K.~Kajantie, K.~Rummukainen
and M.~Shaposhnikov, preprint CERN-TH.6901/93.
\item
\label{oneovern}
V. Jain, Nucl. Phys. B394 (1993) 707 and Max-Planck-Institut preprint
MPI-Ph/92-72; V.~Jain and A.~Papadopoulos, Phys. Lett. B303 (1993) 315 and
Berkeley preprint LBL-33833 (1992); M. Carena and C.E.M. Wagner,
Max-Planck-Institut preprint MPI-Ph/92-67.
\item
\label{epsilon}
P. Ginsparg, Nucl. Phys. B170 (1980) 388;
J. March-Russell, Princeton preprint PUPT-92-1328, LBL-32540.
\item
\label{threed}
J.~Lauer, Heidelberg preprint HD-THEP-92-59;
A.~Jakov\'ac and A.~Patk\'os, Bielefeld preprint BI-TP 93/18
\item
\label{wetterich}
N.~Tetradis and C.~Wetterich, preprint DESY 92-093;
M.~Reuter, N.~Tetradis and C.~Wetterich, preprint DESY 93-004.
\end{enumerate}
\newpage
\section*{Figure captions}
\begin{itemize}
\item[Fig.1:]
The temperature-dependent effective potential of the Standard Model, at
the critical temperature and for $m_h = 60 \gev$, $m_t = 150 \gev$. The
long-dashed line corresponds to the na\"{\i}ve one-loop approximation,
neglecting the scalar loops; the short-dashed line corresponds to the
approximation of eq.~(\ref{carsm}), as in ref.~[\ref{carrington}]; the solid
lines correspond to the approach discussed in the text, for $\gamma =
1,1.5,2,3$.
\item[Fig.2:]
The ratio $v(T_c)/T_c$, as a function of the Higgs mass $m_h$, for
$m_t = 150 \gev$. The long-dashed line corresponds to the one-loop
approximation, the short-dashed line to the approximation of eq.~(\ref{carsm}),
as in ref.~[\ref{carrington}], and the solid lines to the approach discussed in
the text, for $\gamma = 1,1.5,2,2.5,3$.
\end{itemize}
\end{document}